\documentstyle[twocolumn,eqsecnum,aps]{revtex}
\newcommand{\mib}[1]{\mbox{\boldmath $#1$}}

\begin{document}
\draft
\preprint{}
\title{Mixed-Spin Ladders and Plaquette Spin Chains}
\author{ Akihisa Koga, Seiya Kumada, and Norio Kawakami}
\address{Department of Applied Physics,
Osaka University, Suita, Osaka 565, Japan}
\author{ Takahiro Fukui } 
\address{ Institute of Advanced Energy, Kyoto University, 
Uji, Kyoto 611, Japan} 
\date{September 4, 1997}
\maketitle
\begin{abstract}
We investigate low-energy properties of a generalized 
spin ladder model with both of the spin alternation and the bond 
alternation, which allows us to systematically study not only 
ladder systems but also alternating spin chains.
By exploiting non-linear $\sigma$ model 
techniques we study the model with particular
emphasis on the competition between gapful and gapless states.
Our approach turns out to provide a more consistent
semi-classical description of alternating spin chains 
than that in the previous work. We also study a closely related model,
{\it i.e.},  a spin chain with plaquette structure, and show
that frustration causes little effect on its low-energy properties 
so far as the strength of frustration is weaker than a certain 
critical value.
\end{abstract}

\section{Introduction}

Recent extensive experimental and theoretical investigations on spin 
chains and ladders have been providing  a variety of interesting topics.
A typical example is a spin ladder system with and without impurities. 
\cite{RICE,NTAFT,FNST} It has been revealed 
that the impurity effects give rise to drastic changes in the
ladder system, {\it e.g.}, a gapful spin state could be  driven to  
a gapless state, and consequently  lead to a magnetic ordered state. 
Another example is a spin-Peierls system such 
as CuGeO$_3$, for which the effects of frustration and bond alternation,
as well as impurities, play a crucial role.\cite{RRDR,FTS} Also, a 
mixed-spin chain with alternating array of two kind of
spins have stimulated intensive experimental\cite{AltExp} and 
theoretical\cite{AltThe,AltExa,TonHik,Fukui} studies. The alternating
spin chain may be regarded as a high concentration limit of 
magnetic impurities in ordinary spin chains, 
so that this problem has close
relationship to the above-mentioned spin systems with impurities.
A common feature in these problems is how the gapful and gapless 
states compete with each other, providing a variety 
of interesting phenomena.
In this connection, a spin system with plaquette structure
is also interesting,\cite{Plaquette,Ueda,Imada,Ivanov}
for which one can indeed observe how the spin gap is generated
according to a topological nature of the system. 

Motivated by the above stimulating topics, we investigate in this 
paper a quantum spin ladder model with both of the 
spin alternation and the bond alternation.
This system may be referred to as a "mixed-spin ladder".
What is remarkable is that this model allows us to systematically 
study spin ladders, alternating spin chains, and  
spin chains  with periodic array of magnetic impurities.
We take a non-linear $\sigma$ model 
(NL$\sigma$M) approach to describe low-energy 
properties of the model, and  focus particular attention 
on the competition between  gapful and gapless states.
Also, we study a closely related model, {\it i.e.}, 
a spin chain with plaquette structure,\cite{Imada,Ivanov}
which can be naturally constructed from a particular mixed-spin 
ladder model. This model is also instructive to discuss the gap 
formation in quantum  spin systems. 

The paper is organized as follows.
In \S 2, we first introduce the model system, and 
map it to the NL$\sigma$M by taking the continuum limit.
We discuss in \S 3 the properties of two- and three-leg ladder systems
with paying particular attention to the relationship to alternating
spin chains, and then move to many-leg ladder systems 
to discuss spin chains  with a periodic array of magnetic impurities.
In \S 4 we next investigate a plaquette spin chain 
with frustration, and show that frustration little 
affects low-energy  properties when its strength is weaker than 
a certain critical value, being consistent with recent numerical 
findings. Brief summary is given in the last section.

\section{Mapping to the Non-Linear $\sigma$ Model}

We consider a $n_{l}$-leg mixed-spin ladder system with  
the bond alternation, which is described by the following Hamiltonian
\begin{eqnarray}
H&=&\sum_{j=1}^{N}\left\{\sum_{a=1}^{n_{l}}J_{a}\left[ 1+(-1)^{j}
\gamma_{a}\right]{\mib S}_{a,j}\cdot{\mib S}_{a,j+1}\right.\nonumber\\
&&+\left.\sum_{a=1}^{n_{l}-1}J^{\prime}{\mib S}_{a,j}\cdot
{\mib S}_{a+1,j}\right\},
\label{eqn:H1}
\end{eqnarray}
where $N$ is the number of lattice sites for each chain, and 
$S_{a,j}$ is the spin operator at the $j$-th site of the $a$-th chain.
Here  $\gamma_{a}$ is the bond-alternation parameter, and 
$J_{a}$ and $J'$ denote the intraleg and interleg coupling constants,
respectively.  Throughout this paper we assume that the spin 
have different values for different chains, but 
the same value $s_a$ in the same chain. 
In Fig.1, we have drawn the ladder model  schematically.
Since we deal with the case of $J_{a}$, $J^{\prime}>0$, 
and $-1\leq\gamma_{a}\leq1$, the classical minimum 
of the Hamiltonian eq. (\ref{eqn:H1}) is realized in the 
N$\acute{\rm e}$el state. What is most distinct from  
ordinary spin ladders is that the present system includes not 
only the bond alternation, but also the spin alternation. 
This generalization thus allows us to study interesting spin chains
in some limiting cases. For example, setting $J_{a}=0\;(1<a<n_{l})$, 
$\gamma_{1}=1$, and $\gamma_{n_{l}}=-1$,
the model eq. (\ref{eqn:H1}) reproduces an alternating-spin 
chain with the periodic arrangement of spins
$s_{1}\circ s_{2}\circ ...\circ s_{n_{l}}\circ 
s_{n_{l}}\circ ...\circ s_{2}\circ s_{1}$. Moreover, if
we further take a large $n_l$ limit for the ladder composed of a
chain with spin-$s_2$ and all other chains with  spin-$s_1$, the 
Hamiltonian  can be  reduced to the spin-$s_1$ 
chain with a periodic array of dilute spin-$s_2$ impurities.\cite{Fukui}
In this way, by choosing appropriate parameters for the 
number of legs $n_l$, the spin $s_a$ and strength of bond-alternation, 
the mixed-spin ladder model naturally interpolates various 
interesting spin systems which have been intensively studied recently.
We note that a similar but different ladder system
with spin alternation has been discussed  
for the two-leg case.\cite{fuklad}

In the following we shall map the system to the NL$\sigma$M
and discuss its low-energy properties.
To this end, it is convenient to exploit techniques in 
coherent-state path integral formalism. Since the detail of 
the formulation can be found in standard text books\cite{BOOK}
and also in recent papers\cite{fuklad,Sene,Delgado,Aringa},
we briefly summarize how to apply techniques to the present model.
The partition function in this system is given by
\begin{equation}
Z=\int {\rm D}{\mib \Omega}\exp\left[- {\rm i}
\sum_{a,j}s_{a}\omega[{\mib \Omega}_{a}(j)]-
\int_{0}^{\beta} {\rm d}\tau H(\tau )\right],
\end{equation}
where 
\begin{eqnarray}
\omega[{\mib \Omega}]&=&\int_{0}^{\beta} {\rm d}\tau\dot{\varphi}
(1-\cos\theta),\\
H(\tau )&=&\sum_{j=1}^{N}\left\{ \sum_{a=1}^{n_{l}}J_{a}
s_{a}^{2}\left[ 1+(-1)^{j}\gamma_{a}\right]{\mib \Omega}_{a}(j)
\cdot{\mib \Omega}_{a}(j+1)\right.\nonumber\\
&&+\left.\sum_{a=1}^{n_{l}-1}J^{\prime}s_{a}s_{a+1}{\mib \Omega}_{a}(j)
\cdot{\mib \Omega}_{a+1}(j)\right\}.
\label{ham}
\end{eqnarray}
In the above expression we have introduced the unit 
vector ${\mib \Omega}_{a}(j)$ by
${\mib S}_{a,j}=s_{a}{\mib \Omega}_{a}(j)$, where 
${\mib \Omega}=(\sin\theta\cos\varphi,\sin\theta\sin\varphi,\cos\theta)$.
The term $\omega[{\mib \Omega}]$ is the Berry phase which corresponds  
to the solid angle enclosed by the vector 
${\mib \Omega}$ on unit sphere. In our semi-classical approach, 
it is assumed that the spin wave analysis
can correctly describe low-energy modes at wave vectors near $0$ 
and near the ordering wave vector $\pi$.
Then, ${\mib \Omega}_{a}$ may be written as
\begin{eqnarray}
{\mib \Omega}_{a}(j)=(-1)^{a+j}{\mib \phi}_{j}
\left[1-{\mib l}_{a}^{2}(j)\right]^{1/2}+{\mib l}_{a}(j).
\label{order}
\end{eqnarray}
The staggered field ${\mib \phi}_{j}$ in eq. (\ref{order}), 
which corresponds to the component around $\pi$,
is slowly varying field on the scale of lattice spacing.
We have assumed here that the field ${\mib \phi}_{j}$ does 
not depend on the index $a$,
which implies that the spin correlation along rungs is 
sufficiently strong so as to develop the coherence
in this direction; {\it i.e.}, $\xi \gg n_{l}a_{0}$, 
where $\xi$ is the staggered spin correlation length and $a_{0}$ 
is the lattice constant. The other field, ${\mib l}_{a}(j)$, 
is small fluctuation field around $k=0$: $|{\mib l}|\ll 1$.
As is seen below, the introduction of  the $a$-dependence 
for the field ${\mib l}$ improves our semi-classical approximation. 
The constraint ${\mib \Omega}_{a}^{2}(j)=1$ is 
now replaced by ${\mib \phi}_{j}^{2}=1$ and 
${\mib \phi}_{j}\cdot{\mib l}_{a}(j)=0$.
Substituting eq. (\ref{order}) into eq. (\ref{ham}) and 
making the expansion up to quadratic
order in ${\mib l}$, ${\mib \phi}^{\prime}$, and $\dot{{\mib \phi}}$,
we obtain the action in the continuum limit
\begin{eqnarray}
S_{\rm H} &=& \frac{ J_{1}s_{1}^{2} }{ 2 } \int {\rm d}x \int_{0}^{\beta}  
{\rm d}\tau \left\{ K \right.{\mib \phi}^{\prime 2} \nonumber\\
& & + \sum_{a,b} {\mib l}_{a} L_{ab} {\mib l}_{b}+\sum_{a} \left. g_{a}
{\mib l}_{a} \cdot {\mib \phi}^{\prime} 
\right\},
\label{hami}
\end{eqnarray}
where the lattice constant is taken as unity.
For later convenience, we have introduced the parameters 
$K$, ${\mib g}$, and $L$ which are given in terms of 
microscopic parameters in our model.
Their explicit forms will be given in the following sections.
We next take the continuum limit of the Berry phase term,
which results in  an ordinary form,
\begin{eqnarray}
S_{\rm B}&=&{\rm i}\sum_{a,j}s_{a}\omega 
[{\mib \Omega}_{a}(j)] \nonumber \\
&=&-{\rm i}\theta^{\prime}Q - {\rm i}s_{1}\int {\rm d}x\int_{0}^{\beta} {\rm d}\tau\sum_{a}f_{a}
{\mib l}_{a}\cdot{\mib \phi}\times\dot{{\mib \phi}},
\label{berry}
\end{eqnarray}
where $\theta^{\prime}=2\pi\sum_{a}^{n_{l}}(-1)^{a}s_{a}$,
${\mib f}^{t}=(1,\alpha_{2},\alpha_{3},...)$, and
$\alpha_{m}=s_{m}/s_{1}$.
The value $Q$ in eq. (\ref{berry}) is the winding 
number defined by $Q=\frac{1}{4\pi}\int {\rm d}x\int {\rm d}\tau{\mib \phi}
\cdot({\mib \phi}^{\prime}\times\dot{{\mib \phi}})$.
Integrating eqs. (\ref{hami}) and (\ref{berry}) over the fluctuation 
fields ${\mib l}$, we thus end up with the NL$\sigma$M 
with topological term,
\begin{equation}
Z=\int {\rm D}{\mib \phi}\exp\left[{\rm i}\theta Q-\frac{1}{2g}\int {\rm d}x
\int_{0}^{\beta} {\rm d}\tau\left( v_{\rm s}{\mib \phi}^{\prime 2}
+\frac{1}{v_{\rm s}}\dot{{\mib \phi}}^{2}\right)\right],\label{sgm}
\end{equation}
where 
\begin{eqnarray}
\theta&=&\theta^{\prime}+2\pi s_{1}{\mib g}^{t}L^{-1}{\mib f},\nonumber\\
g&=&\frac{1}{s_{1}}\left\{\left(K-\frac{1}{4}{\mib g}^{t}L^{-1}
{\mib g}\right){\mib f}^{t}L^{-1}{\mib f}\right\}^{-1/2},\nonumber\\
v_{\rm s}&=&J_{1}s_{1}\left\{\frac{K-\frac{1}{4}{\mib g}^{t}L^{-1}{\mib g}}
{{\mib f}^{t}L^{-1}{\mib f}}\right\}^{1/2}.
\end{eqnarray}
This completes the basic formulation in our semi-classical
approach.  Note that the topological term in eq.(\ref{sgm})
is quite  essential to classify the behavior of the system;
It is known that the system with $\theta=\pi ({\rm mod} 2\pi)$ 
is gapless, whereas the system with $\theta \neq \pi 
({\rm mod} 2\pi)$ is gapful. Indeed in our analysis, this topological
term plays a central role to specify whether the system is 
gapful or gapless. In the following 
sections, we deal with some interesting examples of 
the model,  and discuss their low-energy properties
based on the NL$\sigma$M approach.

\section{Spin Ladders and Alternating Spin Chains}
\subsection{Two-leg ladders}

Let us start with a two-leg ladder system. 
For $n_{l}=2$, we have the parameters for the NL$\sigma$M,
$K=1+\alpha_{2}^{2}R_{2}$, 
${\mib g}^{t}=(4\gamma_{1}, -4\alpha_{2}^{2}R_{2}\gamma_{2})$, 
${\mib f}^{t}=(1, \alpha_{2})$, and
\begin{equation}
L=\left(
\begin{array}{cc}
4+\alpha_{2}R^{\prime}&\alpha_{2}R^{\prime}
\\\alpha_{2}R^{\prime}&4\alpha_{2}^{2}R_{2}+\alpha_{2} R^{\prime}
\end{array}\right),
\end{equation}
where $R_{2}=J_{2}/J_{1}$ and $R'=J^{\prime}/J_{1}$.
>From these expressions, we obtain
\begin{eqnarray}
\theta&=&2\pi (s_{2}-s_{1})+\frac{2\pi s_{1}}{R^{\prime}+4\alpha_{2}R_{2}+\alpha_{2}^{2}R^{\prime}R_{2}}\nonumber\\
&\times&\left\{R^{\prime}(1-\alpha_{2})(\gamma_{1}+\alpha_{2}^{2}
\gamma_{2}R_{2})+4\alpha_{2}R_{2}(\gamma_{1}-\alpha_{2}\gamma_{2})\right\}.\nonumber\\
\end{eqnarray}
In the case of $\gamma_1=\gamma_2=0$, the system is reduced 
to a ladder system without bond alternation,
for which the topological term is given by
$\theta=2 \pi (s_2-s_1)$.\cite{Sene}   Therefore, if  either of the 
spin in the two chains has a half-integer spin, 
the system becomes gapless, otherwise the system
is gapful. If we include the bond alternation, the system is 
gapful in general.

Since essential properties for ordinary two-leg ladder systems
have been already clarified, we focus
our attention on some characteristic  cases,
 and confirm that our results consistently reproduce those 
derived previously.\cite{Fukui,Delgado}
Let us first consider the uniform spin case
 $s_{1}=s_{2}\equiv s$, with  a special bond-alternation
$\gamma_{1}=-\gamma_{2}=\gamma$, 
$R^{\prime}=R$, and $R_{2}=1$.  Then the system
is reduced to an ordinary spin chain 
with the bond alternation (its structure is similar   
to a "snake chain"). 
In this case, we obtain the 
topological term  with $\theta =8\pi s\gamma /(2+R)$, which 
agrees with those derived by Delgado {\it et al}.\cite{Delgado}  
Another interesting limit is the
case of $\gamma_{1}=-\gamma_{2}=1$, $R^{\prime}=2$, $R_{2}=1$, 
and $J\rightarrow J/2$. In this case also, the 
system is reduced to a snake chain, but 
it becomes a mixed-spin chain with  alternating 
array of spins,\cite{TonHik,Fukui}
$s_{1}\circ s_{1}\circ s_{2}\circ s_{2}\circ s_{1}\circ s_{1}
\circ s_{2}\circ s_{2}\circ ...$. Here we have to change the lattice 
constant $1\rightarrow 2$ to obtain the straight chain from 
the snake chain. Defining the effective 
spin $s^{\prime}=2s_{1}s_{2}/(s_{1}+s_{2})$, 
we have  $\theta=2\pi s^{\prime}$, $g=2/s^{\prime}$, 
and $v_{\rm s}=2Js^{\prime}$.  Therefore, the system is 
gapful in general except for special cases 
for which $\theta= \pi $. It is seen that 
these results agree with those obtained 
directly from  the single chain model.\cite{Fukui}  
The coincidence is apparent since in the two-leg ladder system the 
way how the fluctuation ${\mib l}_{a}(j)$ is introduced is the same as 
in the previous one\cite{Fukui}. In the following subsection, however, 
we shall see that our approach based on a generalized ladder model
indeed improves a semi-classical description of  alternating
spin chains.

\subsection{Three-leg ladders}
We now turn to a three-leg ladder system.
For $n_{l}=3$ the parameters for the NL$\sigma$M
read $K=1+\alpha_{2}^{2}R_{2}+\alpha_{3}^{2}R_{3}, \theta^{\prime}=-2\pi (s_{1}-s_{2}+s_{3})$, 
${\mib g}^{t}=(4\gamma_{1}, -4\alpha_{2}^{2}R_{2}\gamma_{2}, 4
\alpha_{3}^{2}R_{3}\gamma_{3})$, ${\mib f}^{t}=(1, \alpha_{2}, 
\alpha_{3})$, and
\begin{equation}
L=\left(
\begin{array}{ccc}
4+\alpha_{2}R^{\prime}&\alpha_{2}R^{\prime}&0\\
\alpha_{2}R^{\prime}&4\alpha_{2}^{2}R_{2}+\alpha_{2}(1+\alpha_{3})R^{\prime}&
\alpha_{2}\alpha_{3}R^{\prime}\\
0&\alpha_{2}\alpha_{3}R^{\prime}&4\alpha_{3}^{2}R_{3}+\alpha_{2}\alpha_{3}R^{\prime}
\end{array}
\right),
\end{equation}
where $R_{3}=J_{3}/J_{1}$.

Although our three-leg ladder model 
includes  various cases according to the choice of the
model parameters, we again focus our discussions on some 
interesting cases.  
We start with a simple case for which
$\gamma_1=\gamma_2=\gamma_3=\gamma$:
three chains with different spins  
are simply connected. In this case, we can typically see how
the topological nature of mixed spins shows up in low-energy physics.
The corresponding topological term is reduced to the form, 
\begin{equation}
\theta=2 \pi (s_{1}-s_{2}+s_{3})(\gamma -1).
\end{equation}
An ordinary three-leg ladder is given by $s_1=s_2=s_3=1/2$,
for which we have $\theta=\pi$ (mod $2\pi$) for $\gamma=0$, and 
hence the system becomes gapless, as is well known. 
For general $\gamma$,
 $\theta$ takes values different from $\pi$ (mod $2\pi$), 
resulting in the gapful phase. 

A remarkable point we wish to mention here is that there is 
a particular combination of spins, $s_1+s_3=s_2$, for 
which the bond-alternation parameter $\gamma$ disappears from the 
expression of the topological term.  
For example, for the simplest 
case with  $s_1=s_3=1/2$, $s_2=1$, which would be   
a possible candidate for experimental realization,
we always have  $\theta=0$ irrespective of the 
bond alternation. This is the remarkable result for which 
the topological nature of spins essentially
determines whether the system is gapful or gapless, 
hiding the effects of the bond alternation.  This implies that
the Berry phase arising from the bond-alternation
cancels each other due to the simple spin configuration
of the system.

The second example we consider here is again a simple  model 
with  $\gamma_{1}=-\gamma_{3}=1$, 
$\gamma_{2}=\gamma$, $R_{3}=1$, $\alpha_{
2}=\alpha,$ and $\alpha_{3}=1$,
which indeed exhibits the interplay of the spin alternation
and the bond alternation.
In particular, in this model, the three-leg ladder system is reduced to 
a kind of plaquette chain 
(Fig.2, see also the next section), 
for which a square plaquette 
is connected to its next plaquette sharing one of its corner 
alternately. 
We shall see that this system still possesses nontrivial interesting 
cases including the alternating 
spin chain. The topological term $\theta$ has the form
\begin{eqnarray}
\theta=2\pi s_{2}-\frac{2\pi s_{1}\alpha^{2}\gamma R_{2}(4-2R^{\prime}+\alpha R^{\prime})}{2R^{\prime}+4\alpha R_{2}+\alpha^{2}R^{\prime}R_{2}}.
\label{theta}
\end{eqnarray}
Let us now discuss whether the system is gapless or gapful
according to the topological term.  
One can easily see from (\ref{theta})
that $\theta$  takes values different from
$\pi$ in general, resulting in a gapful phase.  
We find three possibilities 
to let $\theta$ be equal to $2\pi s_{2}$, for which the system
can be gapless for a half-integer $s_2$:
(i) $R_{2}=0$, (ii) $\gamma=0$, and (iii) 
$R^{\prime}=4/(2-\alpha )$.

In the case of (i) our model is further reduced to the 
alternating spin chain, $s_{1}\circ s_{1}\circ s_{2}\circ 
s_{1}\circ s_{1}\circ s_{2}...$ with singlet ground state.
Setting $R^{\prime}=2$, $J_{1}\rightarrow J_{1}/2$, and the lattice 
constant $1\rightarrow3$, we obtain the parameters  
\begin{eqnarray}
g&=&\frac{2}{s_{1}}\frac{2+\alpha}{\sqrt{\alpha(8+\alpha^{3})}},\\
v_{\rm s}&=&6J_{1}s_{1}\sqrt{\frac{\alpha}{8+\alpha^{3}}}.
\end{eqnarray}
We have numerically checked that the velocity $v_{\rm s}$ 
coincides with the velocity $v_{{\rm sw}}$ which is estimated 
by the spin wave analysis. In contrast to this consistent result, 
when the NL$\sigma$M techniques are naively applied to
a single chain model,\cite{Fukui}  one may encounter
a pathological result for which $v_{\rm s}$  and $v_{\rm sw}$  
have different values.
This implies that the present approach improves the previous 
one, and provides a NL$\sigma$M  mapping consistent with the 
spin wave analysis.

The results in the case (ii) are rather simple, in which
the system is decomposed into two parts:
One is the dimer part which has singlet state of two $s_{1}$ spins, 
and the other is the spin-$s_{2}$ chain. The critical behavior is
thus determined by the $s_2$-spin chain.  We have 
to note, however, that in this case our approximation 
based on a NL$\sigma$M becomes worse, because 
in our approach the spin coherence is assumed to be 
well developed among chains, whereas in this limiting case
such coherence is not developed. So, the present approach may
describe only the qualitative properties in this limiting case.
Lastly, we point out a novel behavior 
observed in the case (iii). It may be rather surprising that 
the topological term in this case is controlled only by
the spin $s_{2}$ in spite of the existence of the 
bond-alternation $\gamma$, which is quite contrasted to  
the behavior observed in 
ordinary spin chains with bond-alternation $\gamma$, where
the topological term should depend on $\gamma$.
This phenomenon may come from the interplay 
between the bond alternation and the spin alternation.
All the above three examples may give characteristic 
phenomena inherent in ladder systems with the 
spin alternation.

\subsection{Spin chains with magnetic impurities }

We have seen so far  that the mixed-ladder model allows us to 
naturally interpolate  the physics of ladders and 
that of spin chains. Stimulated by this, we further extend 
the analysis of the case (i) to a larger $n_l$ case 
to discuss a spin chain with magnetic impurities. 
Although the spin chain with periodic impurities seems a little bit
peculiar at first glance, it still involves some essential properties
expected for ordinary impurities, as claimed in refs.
\cite{Fukui,iino}

We start with the multiple ladder Hamiltonian (\ref{eqn:H1}) 
with $n_{l}=2m+ 1$ 
which has the spin $s_{a}=s_{1}$ $(s_{2})$ for $a\neq m+1$ ($a=m+1$).
Setting $J_{a}=0\;(1<a<n_{l})$, $2J_{1}=2J_{n_{l}}=J^{\prime}=J$, $\gamma_{1}=1$, $\gamma_{n_{l}}=-1$, 
and redefining the lattice constant $1\rightarrow n_{l}$ 
in this model, we arrive at 
the spin-$s_{1}$ chain with periodic array of spin-$s_{2}$
impurities. The case (i) in the previous subsection corresponds
to $m=1$. We analyze this model semiclassically as done before. 
As a result, we arrive at the NL$\sigma$M with
topological term $\theta=2\pi s_{2}$. 
Therefore, we can say that the system may be gapful (gapless), when
$s_2$ is integer (half-integer), as is naturally seen 
>from the topological nature of the system.\cite{Fukui}
This implies that well-separated $s_2$ spins
correlate with each other making a narrow gapless band,
even if the background spin $s_1$ forms the Haldane gap. 

As mentioned above, our approach yields qualitatively correct 
results even for the case of dilute impurities.
Now, the question is how consistent  the present NL$\sigma$M
approach is. For this purpose we have numerically checked again that the 
velocity $v_{{\rm s}}$ in the $m\rightarrow\infty$ limit agrees 
with $v_{{\rm sw}}$ deduced from  the spin-wave spectrum,
although the ways to obtain the spin velocities
seem quite different in two formulations.
This implies that our treatment based on the NL$\sigma$M
is, at least, a consistent semiclassical approach even for
the dilute-impurity case.  As mentioned in the previous 
subsection, this indeed improves the previous results obtained
directly from the spin chain model.\cite{Fukui}
The reason why our treatment based on the ladder model
has such advantage is as follows: 
In the ladder system the correlation effects between 
two adjacent $s_2$ spins in the same chain can be 
naturally taken into account rather well.
These two $s_2$ spins  are then reduced 
to two well-separated magnetic impurities in the    
spin chain model, for which the interference between 
them is quite essential for low-energy properties.\cite{velocity}
If we start the single chain model,\cite{Fukui} it is not easy to
incorporate the correlation between largely separated spins.
We would hence say that our 
approach based on the ladder system may provide a more 
efficient framework to
incorporate such interference among magnetic impurities.
It should be noted, however, that in order to discuss
low-energy properties more quantitatively, it is necessary 
 to integrate out the gapful degrees of freedom
properly ($s_1$-spin sectors). This point should be further 
improved by taking into account the quantum fluctuations more
precisely. We believe that the method proposed recently\cite{nagaosa}
should provide an efficient way to resolve this problem.


\section{Plaquette Spin Chains}

We have discussed so far how the gapful and gapless
states compete with each other reflecting 
the topological nature of the system, the interaction strength, etc.
In this connection, we now wish to discuss a 
closely related model, {\it i.e.},
a spin chain model with plaquette structure, which 
is also instructive to discuss how the 
spin gap formation occurs in quantum spin systems.
As far as the spin gap formation is concerned, our system 
may be related to two-dimensional plaquette spin systems 
studied extensively.\cite{Plaquette,Ueda,Imada}  
We show that the universality class of the plaquette chain
is the same as that of a  mixed-spin ladder.

The plaquette spin chain can be naturally constructed
>from  a three-leg ladder system. 
To see the way  clearly, we start by observing how  a 
$n_{l}$-leg ladder system with uniform spin $s$,
for which exchange coupling along the rungs is  
{\it ferromagnetic} ({\it i.e.}
we choose $J' < 0$ and  $J>0$), 
can be unified into an effective single chain. 
As is well known,\cite{ferro} 
low-energy properties of the system is
identical with those for the single  spin-$n_{l}s$ chain. 
This can be explicitly shown by the NL$\sigma$M approach, 
and we arrive at the NL$\sigma$M with the effective intrachain 
coupling $J/n_{l}$. The resulting
 velocity $v_{{\rm s}}=2Js$ again turns out to be  
the same as $v_{{\rm sw}}$ obtained by the spin-wave 
analysis. It is to be noted 
 that the value of $J^{\prime}$ does not appear 
in the parameters in the effective NL$\sigma$M within 
the present approximation.

Based on the above observation, we now consider a specific spin ladder
system composed of three chains with 
spins $s$, $2s$, and $s$, respectively (see Fig.3).
Applying the idea outlined above, 
let us decompose the middle spin-$2s$ chain into two spin-$s$ chains 
with ferromagnetic interchain couplings. The effective
four-chain system is now considered in the continuum limit.
In order to map the system 
to the NL$\sigma$M, we introduce the sigma-model field as
\begin{eqnarray}
{\mib\Omega}_{a}(j)=(-1)^{n_{a}+j}{\mib\phi}_{j}\left[ 
1-{\mib l}_{a}^{2}(j)\right]^{-1/2}+{\mib l}_{a}(j)
\end{eqnarray}
where $n_{a}=1 (0)$ for $a=1, 4 (a=2, 3)$.
The resulting NL$\sigma$M has following parameters
\begin{eqnarray}
K=2+4R_{2},
\end{eqnarray}
\begin{equation}
L=\left(
\begin{array}{cccc}
4+2R^{\prime}&R^{\prime}&R^{\prime}&0\\
R^{\prime}&8R_{2}+ 2R^{\prime}- R_{3}&R_{3}&R^{\prime}\\
R^{\prime}&R_{3}&8R_{2}+ 2R^{\prime} - R_{3}&R^{\prime}\\
0&R^{\prime}&R^{\prime}&4+2R^{\prime}
\end{array}
\right),
\end{equation} 
\begin{equation}
{\mib g}=4\gamma\left(
\begin{array}{c}
1\\0\\0\\-1
\end{array}
\right) ,{\mib f}=\left(
\begin{array}{c}
1\\1\\1\\1
\end{array}
\right).
\end{equation}

Now, setting $\gamma=1$, $R_{2}
= 0$, and redefining the lattice 
constant $1\rightarrow 3$ and the coupling constant $J_{1}\rightarrow J_{1}/2$, we end up with the 
plaquette chain schematically shown 
in Fig.3. 
The corresponding NL$\sigma$M has the 
$\theta =0$, $g=\sqrt{1+R^{\prime}}/s$, and 
$v_{\rm s}=3J^{\prime}s/\sqrt{1+R^{\prime}}$.
In this way, low-energy properties of our plaquette spin chain is 
naturally related to those of a mixed-spin ladder
discussed so far.

It is to be noticed  that $\theta$ is always zero,
which implies that the plaquette spin system is 
in the gapful phase  irrespective 
of the model parameters.  Another remarkable point is that 
 $R_{3}$ does not enter in these parameters 
whether it is positive or negative.
When $R_{3}$ is negative, namely, in the case of 
ferromagnetic coupling $J_{3}$, the resulting gapful phase 
is naturally understood from the above discussions 
for the three-leg chain with spins $s$, $2s$, and $s$.
Namely,  our plaquette spin chain belongs 
to the same universality class of the above  
three-leg ladder system with spins $s$, $2s$ and $s$. 

On the other hand, in the case of antiferromagnetic 
coupling  $R_{3}>0$, the result is nontrivial, because the
system is now subject to frustration in the plaquette. 
By using the case of $R_3>0$, we in turn have an opportunity
to discuss the effects of frustration on our plaquette chain.
For this purpose, we first note that the above gapful phase should be 
changed to another phase when the strength of frustration increases.
Then, the question is to what extent this gapful phase 
is stable against the frustration. To check this point, we here 
recall that the spin wave 
spectrum should have a linear dispersion in order for the system to
be mapped to the NL$\sigma$M.
Keeping this fact in mind, we 
reexamine the spin wave spectrum for the present system.
The behavior of the spin wave dispersion for various values of 
$R_{3}$ is shown 
in Fig.4.
There are four modes, one of which indeed shows 
a linear dispersion  in the low-energy regime.
Let us now focus on the dispersionless mode,
which is decoupled from other collective excitations and depends only on the coupling $J_{3}$.
It may contain key information about the stability
of the system. Increasing the coupling $R_3$ from the ferromagnetic 
to anti-ferromagnetic regime, this mode goes downward
uniformly, and finally reaches  the $p$ axis when $R_{3}=R^{\prime}$.
Beyond this critical value, $R^{\prime}$, the above spin wave
excitations are not well-defined. Therefore we find that our 
NL$\sigma$M analysis of the plaquette spin chain 
holds valid only in the region $R_{3} < R^{\prime}$ where 
the mapping to the NL$\sigma$M is allowed.

A remarkable point we wish to stress is that the 
lowest spin-wave mode is not influenced 
by the change of the coupling strength, implying 
that $v_{{\rm sw}}$ does not change.  This is consistent
with the above results for the NL$\sigma$M
that $R_3$ does  not enter in the model parameters.
Summarizing the above facts, we come to the following conclusion:
even if the coupling $R_3$ in the plaquette is increased,
the effect of frustration little affects the low-energy properties 
of the model in the region for $R_{3}< R'$, at which
the system may undergo a phase transition. We find  this 
remarkable result to be consistent 
with the recent numerical studies on the plaquette 
spin chain.\cite{Ivanov} 
Although in the present approach, we cannot say what happens beyond
the critical $R'$, the numerical study \cite{Ivanov}
has predicted that the 
system should undergo a first-order phase transition to
enter another gapful phase.

\section{Summary}

We have investigated low-energy properties of a spin ladder system
with both of the spin alternation and the bond alternation, which 
has been shown to involve various interesting spin systems.
In particular, we have discussed how the gapless and
gapful states compete with each other
according to the topological nature  of the system.
Starting with the spin-wave analysis on the 
spin ladder system, and introducing fluctuation fields, 
we have mapped the system to the NL$\sigma$M.
By using the mixed-spin ladder systems composed of multiple 
chains, we have discussed characteristic properties of
the alternating spin chain  
as well as the spin chain with magnetic impurities.
We have found that our approach in the present study provide a more 
consistent semi-classical description for such spin chains
than the previous work based on the single-chain model. 

We have also studied the plaquette spin chain, and shown that
the system belongs to the same universality class of 
the three-leg ladder system composed of
chains with spins $s$, $2s$ and $s$. The effects
of frustration on the plaquette chain has then  been investigated.
It has been shown that the frustration little affects low-energy
properties when its strength is weaker than a certain 
critical value.

In this paper, we have restricted our discussions 
to several specific spin models to demonstrate how 
the spin alternation and the bond alternation 
affects the low-energy properties.  By extending and improving our 
treatments, we think that various interesting 
quantum spin systems can be described systematically in the same
framework of the mixed-spin ladder system, which should be
done in the future study.

\section*{Acknowledgements}
The work is partly supported by a Grant-in-Aid from the Ministry of 
Education, Science, Sports and Culture.



\begin{figure}[h]
\vspace{0cm}
\caption{
Mixed-spin ladder with the bond alternation.}
\label{fig:ladder}
\end{figure}

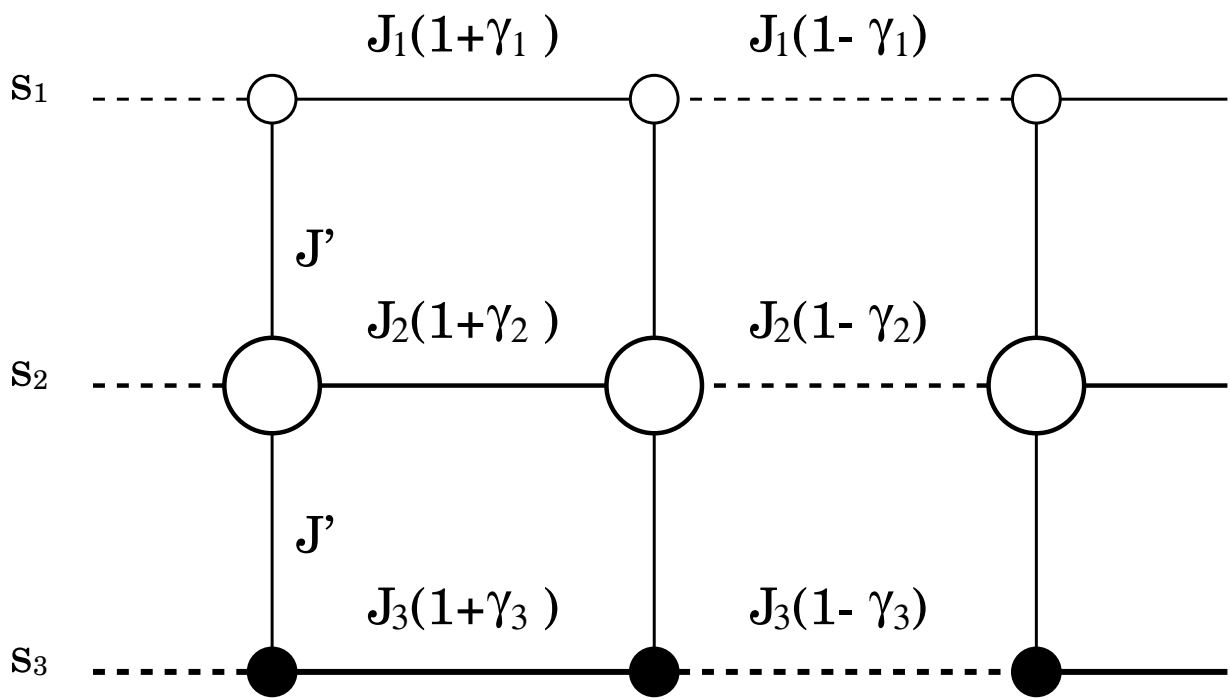
\begin{figure}[h]
\vspace{0cm}
\caption{
A special model for the three-leg ladder system.
}
\label{fig:three}
\end{figure}

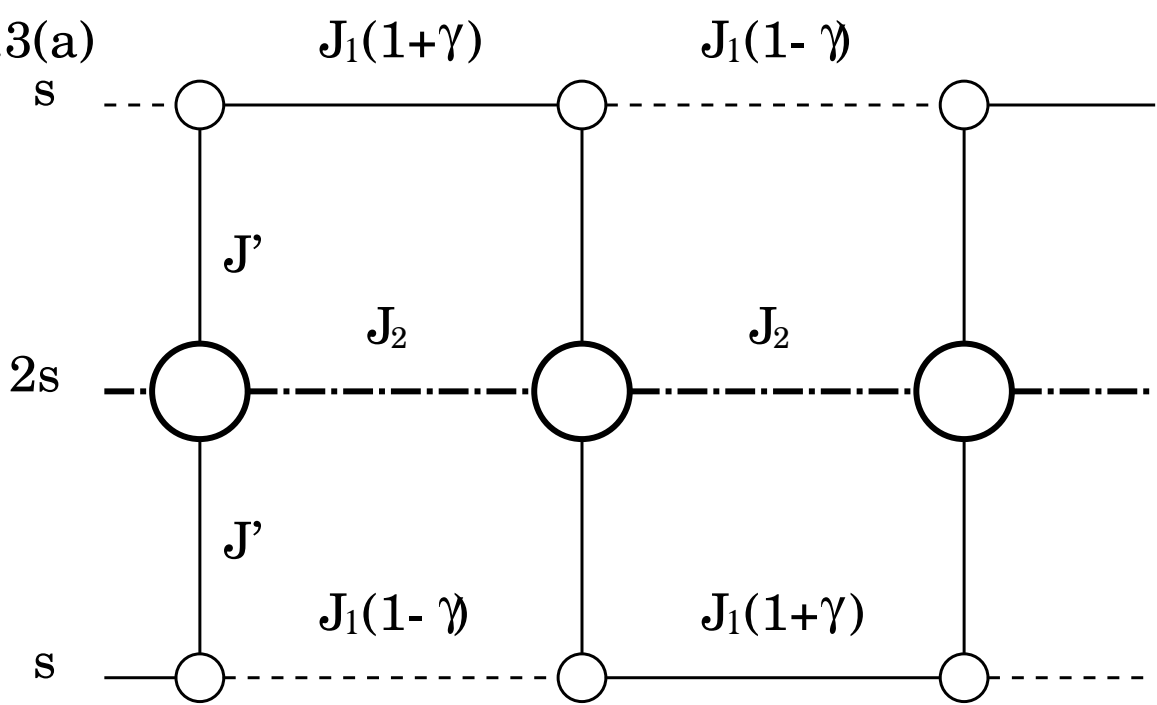
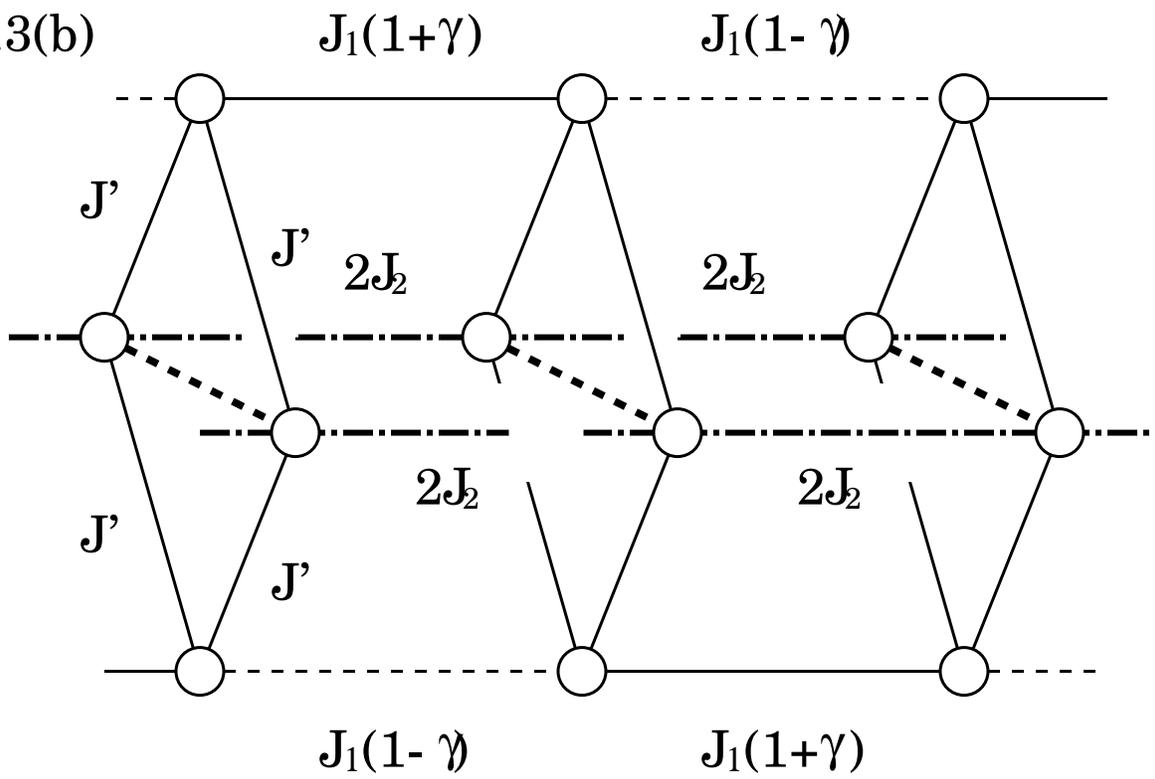
\begin{figure}[h]
\vspace{0cm}
\caption{
(a)Spin ladder composed of three chains with spins $s,2s$, and $s$. 
(b)Corresponding four chain system (See the text). 
(c)Spin chain with plaquette structure. 
In the system (b) by choosing $J_{2}=0 $ and $ \gamma=1$, and redefining the coupling constant $J_{1}\rightarrow J_{1}/2$, we have the plaquette spin chain
(c).
}
\label{fig:plaquette}
\end{figure}

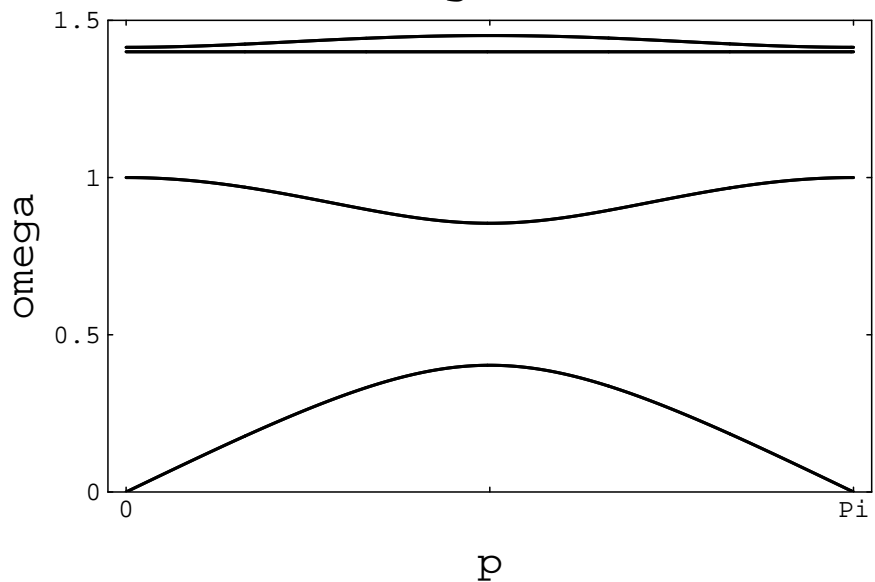
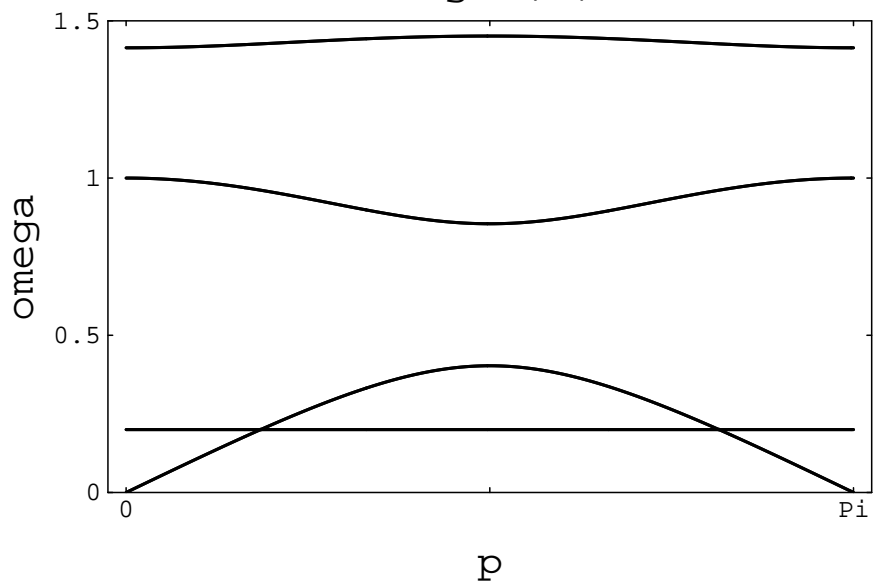
\begin{figure}[h]
\vspace{0cm}
\caption{
Spin-wave spectrum as functions of momentum $p$ for $J_{3}=-0.4$(a), $0.8$(b), and $1.0$(c) with other parameters being fixed as $s=1/2, J=J^{\prime}=1, \gamma=1$, and $J_{2}=0$.
}
\label{fig:spinwave}
\end{figure}


\begin{thebibliography}{99}
\bibitem{velocity}
Note that the caluculated spin velocity  for 
$m \rightarrow \infty$ is less than
that expected for the pure case;  $v_{{\rm s}} < 2Js_{1}$.
This result implies that our approach indeed 
incorporate interference among impurities even for the 
dilute limit.
\end{thebibliography}
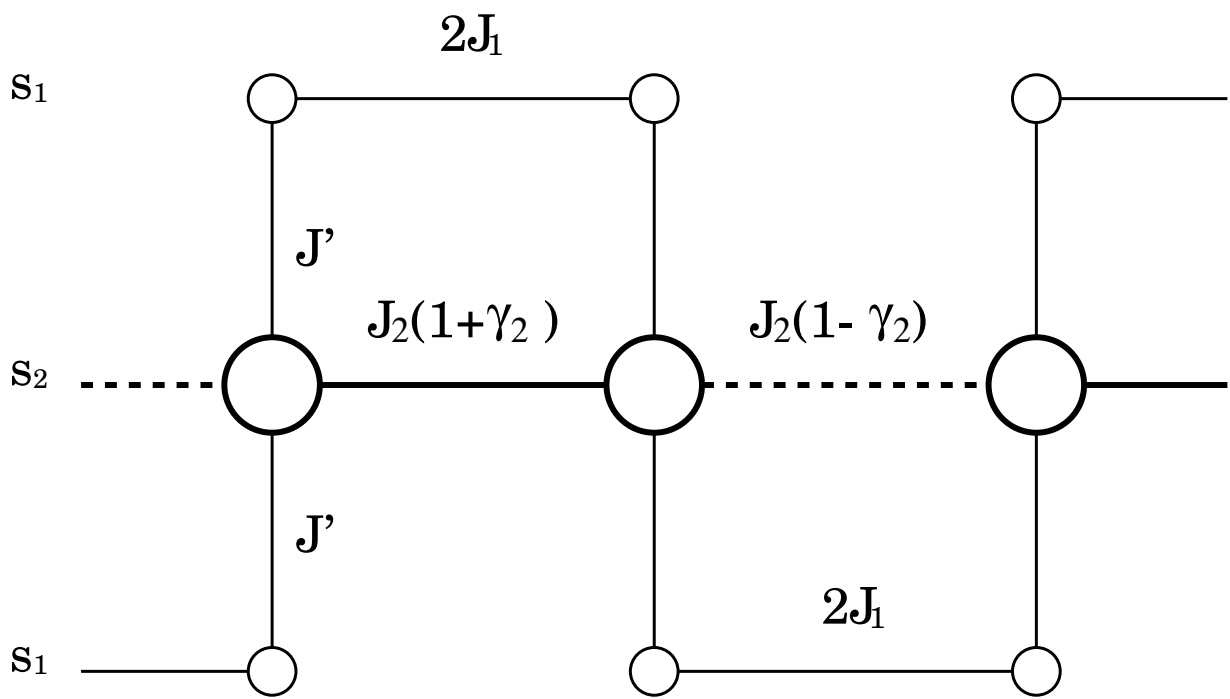
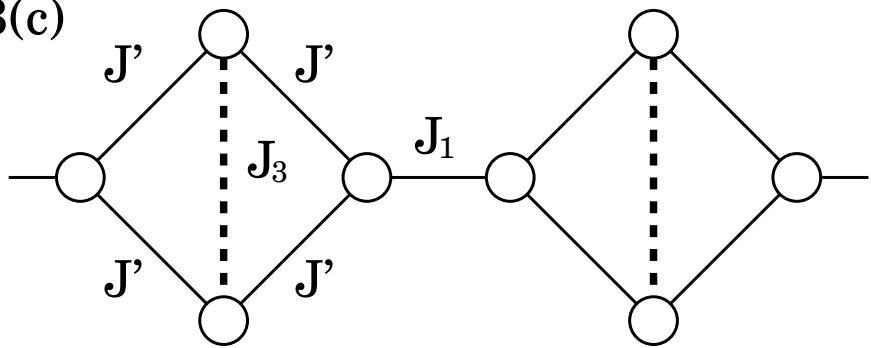
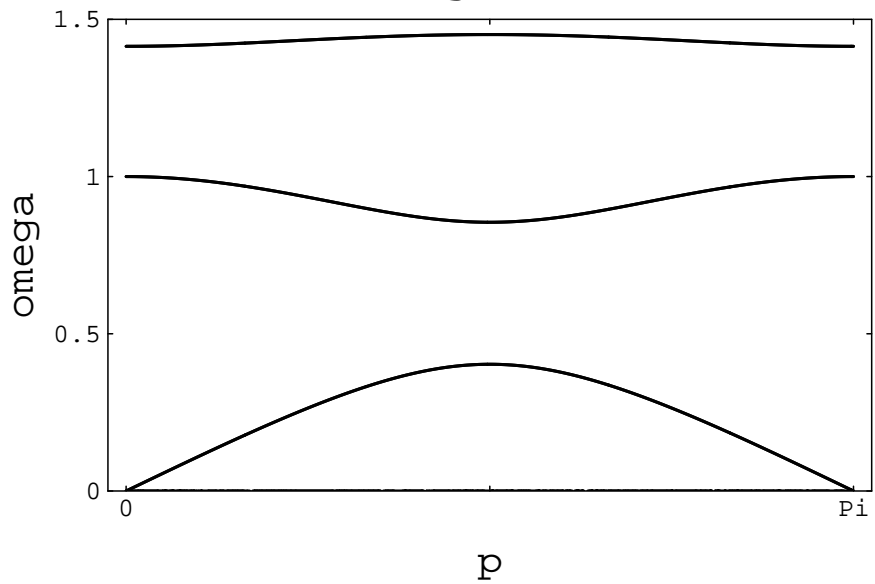
\end{document}